\documentclass[preprint]{aastex}

\received{1 Sep 2004}
\revised{4 Mar 2005 and 4 Jun 2005} 
\accepted{}

\shortauthors{Fynbo et al.}
\shorttitle{On the Afterglow and Host Galaxy of GRB~021004}

\slugcomment{Resubmitted to ApJ March 4 2005 and again June 5 2005}

\begin{document}

\title{On the Afterglow and Host Galaxy of GRB~021004: A Comprehensive 
Study with the \textit{Hubble Space Telescope}\footnote{Based on observations made 
with the NASA/ESA {\it Hubble
Space Telescope}, obtained at the Space Telescope Science Institute,
which is operated by the Association of Universities for Research in
Astronomy, Inc., under NASA contract NAS 5-26555. These observations
are associated with programs 9074 and 9405.}}

\author{
J.~P.~U. Fynbo\altaffilmark{2},
J.~Gorosabel\altaffilmark{3,4},
A.~Smette\altaffilmark{5,6},
A.~Fruchter\altaffilmark{3},
J.~Hjorth\altaffilmark{2},
K.~Pedersen\altaffilmark{2},
A.~Levan\altaffilmark{8},
I.~Burud\altaffilmark{3},
K.~Sahu\altaffilmark{3},
P.~M.~Vreeswijk\altaffilmark{5},
E.~Bergeron\altaffilmark{3},
C.~Kouveliotou\altaffilmark{12},
N.~Tanvir\altaffilmark{9},
S.~E.~Thorsett\altaffilmark{11},
R.~A.~M.~J.~Wijers\altaffilmark{16},
J.~M.~Castro Cer\'on\altaffilmark{2,3},
A.~Castro-Tirado\altaffilmark{4},
P.~Garnavich\altaffilmark{13},
S.~T.~Holland\altaffilmark{14,15,16},
P.~Jakobsson\altaffilmark{2},
P.~M\o ller\altaffilmark{7},
P.~Nugent\altaffilmark{10},
E.~Pian\altaffilmark{17},
J.~Rhoads\altaffilmark{3},
B.~Thomsen\altaffilmark{18},
D.~Watson\altaffilmark{2},
and
S.~Woosley\altaffilmark{11}
}

\altaffiltext{2}{Niels Bohr Institute, University of Copenhagen,
Juliane Maries Vej 30, DK--2100 Copenhagen~\O, Denmark}
\altaffiltext{3}{Space Telescope Science Institute, 3700 San Martin
Drive, Baltimore MD 21218, USA}
\altaffiltext{4}{Instituto de Astrof\'\i sica de Andaluc\'\i a (IAA-CSIC),
Apartado de Correos, 3004, E-11080, Granada, Spain}
\altaffiltext{5}{European Southern Observatory, Casilla 19001, Santiago 19,
Chile }
\altaffiltext{6}{Research Associate, National Fund for Scientific Research 
(FNRS), Belgium}
\altaffiltext{7}{European Southern Observatory, Karl Scwarzschild-Strasse 2,
D-85748 Garching bei M{\"u}nchen, Germany}
\altaffiltext{8}{Department of Physics and Astronomy, University of
Leicester, University Road, Leicester, LE1 7RH, UK}
\altaffiltext{9}{Department of Physical Science, University of
Hertfordshire, College Lane, Hatfield, AL10 9AB, UK}
\altaffiltext{10}{Lawrence Berkeley National Laboratory, 1 Cyclotron Road,
Berkeley CA 94720, USA }
\altaffiltext{11}{Department of Astronomy and Astrophysics,
University of California, 1156 High Street, Santa Cruz CA 95064, USA}
\altaffiltext{12}{NASA/Marshall Space Flight Center, National Space Science
and Technology Center (NSSTC), SD-50, 320 Sparkman Drive, Huntsville AL
35805, USA}
\altaffiltext{13}{Astronomical Institute, University of Amsterdam,
Kruislann 403, 1098 SJ Amsterdam, The Netherlands}
\altaffiltext{14}{Department of Physics, University of Notre Dame, 225
Nieuwland Science Hall, Notre Dame IN 46556, USA}
\altaffiltext{15}{Swift Science Center, Goddard Space Flight Center,
Code 660,1, Greenbelt MD 20771-0003, USA}
\altaffiltext{16}{Universities Space Research Association}
\altaffiltext{17}{Osservatorio Astronomico di Trieste, Via
G.B. Tiepolo 11, 34131 Trieste, Italy}
\altaffiltext{18}{Department of Physics and Astronomy, University of Aarhus,
Ny Munkegade, DK-8000 Aarhus C}

\begin{abstract}
We report on Hubble Space Telescope (\textit{HST}) observations of the 
late-time afterglow and host galaxy of GRB~021004 ($z = 2.33$). Although 
this gamma-ray burst (GRB) 
is one of the best observed so far in terms of sampling in the 
time domain, multi-wavelength coverage and polarimetric observations,
there is substantial disagreement between different interpretations 
of data sets on this burst in the literature. We have observed the field of 
GRB~021004 with the \textit{HST} at multiple epochs from 3 days until almost 
10 months after the burst. With STIS prism and G430L
spectroscopy we cover the spectral region from about 2000 \AA \ to 
5700 \AA \
corresponding to 600--1700 \AA \ in the restframe. From the limit on the 
flux recovery bluewards of the Lyman-limit we constrain the 
\ion{H}{1} column density to be above $1\times10^{18}$ cm$^{-2}$ (5$\sigma$). 
Based on ACS and NICMOS imaging we find that the afterglow evolved 
achromatically within the errors (any variation must be less then 5\%)
during the period of \textit{HST} observations. The color changes observed 
by other authors during the first four days must be related to a 
stochastic phenomenon 
superimposed on an afterglow component with a constant spectral shape.
This achromaticity implies that the cooling break has remained on the 
blue side of the 
optical part of the spectrum for at least two weeks after the explosion. 
The optical--to--X-ray slope $\beta_\mathrm{OX}$ is consistent with being the
same at 1.4 and 52.4 days after the burst. This indicates that the
cooling frequency is constant and hence, according to fireball models,
that the circumburst medium has a constant density profile.
The late-time slope of the lightcurve ($\alpha_2$, 
$F_{\nu} \propto t^{-\alpha_2}$) is in the range $\alpha_2 = 1.8$--$1.9$, 
and is inconsistent with a single power-law.
This could be due to a late-time flattening caused by the transition to 
non-relativistic expansion or due to excess emission (a `bump' in the 
lightcurve) about 7 days after burst. The host galaxy is like most previously 
studied
GRB hosts a (very) blue starburst galaxy with no evidence for dust and with 
strong Ly$\alpha$ emission. The star-formation rate of the host is about 
10 M$_{\sun}$ yr$^{-1}$ based on both the strength of the UV continuum and 
on the Ly$\alpha$ luminosity. The spectral energy distribution of the
host implies an age in the range 30--100 Myr for the dominant stellar
population. The afterglow was located very close ($\sim$100 pc) to the center
of the host implying that the progenitor was possibly associated with a 
circumnuclear starburst.
\end{abstract}

\keywords{
cosmology: observations ---
gamma rays: bursts
}

\section{Introduction}

The very rapid localization of cosmic Gamma-Ray Bursts (GRBs) made
possible by the HETE-2 satellite has enabled very
well-sampled multi-band lightcurves ranging from a few minutes
to several months after the bursts. One of the best
studied GRBs so far is the bright burst detected on 2002 October 4 
(all epochs are given as UT) with HETE-2 (Shirasaki et al.\ 2002). 
The optical afterglow was detected unusually early -- 3.2 minutes after
the high energy event (Fox et al.\ 2003) and monitored intensively
with many telescopes in the following hours to months (e.g., Pandey et al.\
2003; Holland et al.\ 2003 -- hereafter H03; Bersier et al.\ 2003; 
Mirabal et al.\ 2003 -- hereafter M03). The degree
of polarization was measured on several epochs leading to the 
detection of a variable polarization angle (Rol et al.\ 2003).
The redshift was determined to be $z=2.33$ based on strong
hydrogen and metal absorption lines as well as a Ly$\alpha$ emission
line from the underlying host galaxy (Chornock \& Filippenko 2002; M\o ller
et al.\ 2002; Pandey et al.\ 2003; Matheson et al.\ 2003; M03).
The absorption system associated with GRB~021004 contained
several components covering a velocity range of more than 3000 km 
s$^{-1}$ with clear evidence for line-locking between the components 
(Savaglio et al.\ 2002; M\o ller et al.\ 2002;  Fiore et al.\ 2004)

Despite the intensive coverage of this afterglow there is quite limited 
agreement between the reported
afterglow parameters and their interpretation, i.e.,  the spectral and 
late-time decay 
slopes, wind or ISM circumburst medium, and position of the cooling break
(e.g., Lazzati et al.\ 2002; Pandey et al.\
2003; Li \& Chevalier 2003; Holland et al.\ 2003; Heyl \& Perna 2003; 
Dado et al.\ 2003; Mirabal et al.\ 2003; Rol et al.\ 2003; Nakar et al.\ 
2003; Bj{\"o}rnsson, Gudmundsson \& J{\'o}hannesson 2004). The reasons are 
{\it i)} the very complex lightcurve during the first week and {\it ii)} 
the presence of a relatively bright host galaxy affecting the analysis of 
the late (fainter) part of the afterglow lightcurve. 

In this paper we present an analysis of \textit{HST} observations 
of the afterglow of GRB~021004 ranging from three days to ten months after the 
burst. Our emphasis is on the early UV spectroscopy, the late-time
afterglow and on the properties of the host galaxy. In Sect.~\ref{sec:obs} we 
describe the observations, data reduction, and analysis and in 
Sect.~\ref{sec:discuss} we present our discussion and conclusions.
We assume $H_0$ = 70 km s$^{-1}$ Mpc$^{-1}$, $\Omega_m$ = 0.3 and 
$\Omega_\Lambda$ = 0.7 throughout.

\section{Observations and Data Reduction}
\label{sec:obs}
 
\textit{HST} observed GRB~021004 after a fast turnaround on 2002 
October 6--7. On 2002 October 6 we used the Space Telescope Imaging 
Spectrograph (STIS) to obtain near-UV spectroscopic observations of the 
afterglow. The Prism (4 orbits) as well as the G430L grating (2 orbits) 
were used for the spectroscopic STIS observations. The G430L spectra
were obtained with a 0.5 arcsec slit and a position angle of 113.8$^o$. 
The observing
log is given in Table~\ref{tab:journal}. The choice of the Prism
could appear surprising at first sight as it provides little coverage
redwards of the Lyman edge at the GRB redshift. However, at the time
of submitting the Phase II proposal, the redshift was 
expected to be $1.60<z<2.1$ based on the presence of a $z=1.60$ \ion{Mg}{2} 
system and the lack of strong Ly$\alpha$ absorption 
(Fox et al.\ 2002; Eracleous et al.\ 2002; Weidinger et al.\ 2002). The 
correct redshift $z = 2.33$ was only announced on 2002 October 8 
(Chornock \& Filippenko 2002)
after the first \textit{HST} observations. The reduction and analysis of 
the spectroscopic observations are described in Sect.~\ref{sec:stis_spectro}.

Following the STIS observation the afterglow was observed with 
the Near Infrared Camera and Multi-Object Spectrometer (NICMOS) in the 
F110W and 
F160W filters and with the Advanced Camera for Surveys (ACS) High Resolution
and Wide Field Cameras (HRC and WFC) in the four filters F250W, F435W,
F606W, and F814W spanning nearly four octaves in wavelength from the near UV 
to the near-IR.
The field was observed again on 2002 October 11, 2002 October 22,
2002 November 26, 2003 May 31, 2003 July 21, and 2003 July 26 in various 
WFC and NICMOS 
filters. The full journal of observations is listed in 
Table~\ref{tab:journal}. The reduction and analysis of the NICMOS and ACS
data are described in Sect.~\ref{sec:imaging}.

\begin{deluxetable}{lccccccc}
\tablecaption{The Log of \textit{HST} STIS, ACS, and NIC3 Observations.
\label{tab:journal}}
\tablehead{
\colhead{Start} &
\colhead{Filter}&
\colhead{Exp.time} & 
\colhead{Aper.} & 
\colhead{Countrate} &
\colhead{AB mag} &
\colhead{AB mag}\\
\colhead{(UT)}&
\colhead{} &
\colhead{(s)} &
\colhead{(arcsec)} &
\colhead{(counts s$^{-1}$)} &
\colhead{(total)} &
\colhead{(OA)}
}
\startdata
STIS \\
\hline
06/10/02  14:34:39 & Prism  & 1300   \\
06/10/02  15:45:23 & Prism  & 3$\times$2625   \\
06/10/02  20:47:37 & G430L   & 900+876 \\
06/10/02  22:09:39 & G430L   & 1200+1261 \\
\hline
ACS \\
\hline
07/10/02  04:42:16 & F250W & 2$\times$2080 & 0.2 & 0.59$\pm$0.04 & 24.18$\pm$0.07 & 24.18$\pm$0.07 \\
07/10/02  11:29:39 & F435W &   520    & 0.5 & 62.56$\pm$0.32 & 21.12$\pm$0.01 & 21.17$\pm$0.01 \\
11/10/02  17:43:38 & F435W & 2$\times$600 & 0.5 & 21.22$\pm$0.20 & 22.29$\pm$0.01 & 22.46$\pm$0.01 \\
26/07/03  02:03:22 & F435W & 4$\times$510 & 0.5 & 2.93$\pm$0.11 & 24.39$\pm$0.04 &    -    \\
07/10/02  11:15:21 & F606W &   520    & 0.5 & 221.29$\pm$0.43 & 20.55$\pm$0.01 & 20.58$\pm$0.01 \\
11/10/02  15:59:24 & F606W & 2$\times$600 & 0.5 & 72.38$\pm$0.34 & 21.77$\pm$0.01 & 21.85$\pm$0.01 \\
22/10/02  18:11:23 & F606W & 4$\times$460 & 0.5/0.2 & 17.43$\pm$0.23 & 23.31$\pm$0.01 & 23.75$\pm$0.01 \\
26/11/02  02:49:17 & F606W & 4$\times$460 & 0.5/0.2 & 7.60$\pm$0.09 & 24.21$\pm$0.01 & 25.68$\pm$0.07 \\
31/05/03  02:36:12 & F606W & 4$\times$480 & 0.5 & 5.56$\pm$0.28 & 24.55$\pm$0.05 &    -    \\
07/10/02  11:01:12 & F814W &   520    & 0.5 & 197.84$\pm$0.35 & 20.09$\pm$0.01 & 20.11$\pm$0.01 \\
11/10/02  16:02:07 & F814W & 2$\times$600 & 0.5 & 61.57$\pm$0.22 & 21.35$\pm$0.01 & 21.42$\pm$0.01 \\
26/07/03  00:12:48 & F814W & 4$\times$480 & 0.5 & 3.76$\pm$0.16 & 24.39$\pm$0.04 &    -    \\
\hline
NIC3 \\
\hline
07/10/02  01:24:05 & F110W & 2$\times$576 & 1.0 & 24.99$\pm$0.75 & 19.86$\pm$0.03 & 19.88$\pm$0.03\\
11/10/02  13:00:20 & F110W & 3$\times$832 & 1.0 & 7.73$\pm$0.25 & 21.14$\pm$0.03 & 21.18$\pm$0.03\\
21/07/03  17:49:34 & F110W & 2$\times$1280 & 0.5 & 0.29$\pm$0.03 & 24.62$\pm$0.10 &    -    \\
07/10/02  03:09:20 & F160W & 576+512    & 0.5 & 31.90$\pm$0.96 & 19.41$\pm$0.03 & 19.43$\pm$0.03\\
11/10/02  14:18:05 & F160W & 3$\times$832 & 1.0 & 9.90$\pm$0.30 & 20.68$\pm$0.03 & 20.74$\pm$0.03\\
22/10/02  16:46:47 & F160W & 3$\times$832 & 1.0 & 2.25$\pm$0.15 & 22.29$\pm$0.08 & 22.57$\pm$0.09\\
25/11/02  07:34:57 & F160W & 3$\times$832 & 0.5 & 0.76$\pm$0.10 & 23.38$\pm$0.13 & 24.67$\pm$0.50\\
26/05/03  00:54:09 & F160W & 2$\times$1280 & 0.5 & 0.47$\pm$0.07 & 23.89$\pm$0.15 &    -    
\enddata
\tablecomments{We here provide
the photometry and the applied apertures sizes. The AB magnitudes are
calculated from the countrates using synphot as described in
Sect.~\ref{sec:photometry}.
The effective central wavelengths of the filters are: 2714 {\AA}
(F250W), 4317 {\AA} (F435W), 5918 {\AA} (F606W), 8060 {\AA}
(F814W), 11229 {\AA} (F110W), and 16034 {\AA} (F160W).}
\end{deluxetable}

\subsection{Spectroscopic STIS Observations}
\label{sec:stis_spectro}

Reduction of the STIS observations were performed using the STIS
Instrument Definition Team version of CALSTIS (Lindler 2003). 

\subsubsection{The Prism Spectra}
\label{sec:prism_spectra}

 The Prism dispersion  is strongly  wavelength dependent (for   further details
see the \textit{HST} Instrument  Handbook, Kim Quijano et al.\ 2003, and Smette
et al.\ 2001).   The dispersion is 40  {\AA}/pixel at 3000 {\AA}, the expected
wavelength of the  Lyman edge at the GRB   redshift and is  smaller at longer
wavelengths. Consequently, a Prism  spectrum is quite sensitive to errors  in
the flat--field,  especially for $\lambda > 3000$~{\AA}.  Also, the  flux
calibration is strongly  dependent on the  accuracy of the zero-point of  the
wavelength calibration. The overall calibration
is unreliable for $\lambda > 3300$ {\AA}.  Over the 300~{\AA}\, useful
range of  our spectra,  these  systematic   effects were  reduced   by
dithering the object along the slit between the four exposures.

In addition, the zero-point of the wavelength calibration also depends
on the precise location  of the object within  the slit. Therefore, we
compared the location  of the afterglow  in the acquisition image with
the location of the slit center. The latter  was measured by fitting a
Gaussian at each row of an image of the slit on  the CCD obtained with
a Tungsten lamp obtained at the end of the first orbit. The centers of
each fitted  Gaussian were themselves  fitted as a function  of column
number with a $3^\mathrm{rd}$ degree polynomial.  At the row where the
afterglow  image reaches a maximum   intensity, the difference between
the afterglow location and the slit center gives an offset of $-0.064$
pixels or  $-0\farcs0016$.   However, an   important  problem is  that
\textit{HST}+STIS shows some  flexure for the  first  few orbits after
acquiring a new target  as the telescope  has changed  its orientation
relative to the sun. Consequently, one cannot be  sure that the object
stayed  over  the same  pixel   during the  4  orbits.  Therefore, the
possibility exists that the  zero point of the wavelength  calibration
estimated during the first  orbit may not be  correct in the following
ones. In order to test and correct for any change in this quantity, we
consider  three different   ways.   We checked  the wavelength  of the
telluric Ly-alpha emission.  However, the width  of the line in pixels
is large because of the large slit width: it is therefore difficult to
measure the its  centroid correctly.  In addition, the  line is at the
opposite part  of the  spectrum relative to  the  interesting data and
therefore our conclusion would be quite sensitive to  a small error in
the  dispersion coefficients.   We also  compared the  location of the
Lyman-break   visible    in  the   individual  spectra     (see  below). 
Unfortunately, the spectrum of  the first orbit  has half the exposure
time of the subsequent ones. In addition, the spectra of the first and
second orbits    show structures    probably  caused   by    incorrect
flat--fielding. We therefore found   that the most reliable  method to
fix the wavelength calibration relative to the one  of the first orbit
is to use the location of the build-up at  the red end of the spectra,
caused by the decreasing dispersion  of the PRISM (D. Lindler, private
communication). The precision of the zero point calibration for the
individual spectra is estimated to be better than 0.3 pixels.

The reduced spectrum is shown in Fig.~\ref{fig:stis_spectra}.  A break
is seen  in the  spectrum at  $\lambda \sim  3000$ {\AA}.  However, an
excess of emission  is seen shortward of it.  We will come back to this
point     in  \S  \ref{sec:lya_138}.      The   first  two   rows   of
Table~\ref{tab:prism} gives the countrate and the integrated flux over
3000  {\AA} $<\lambda<$ 3300 {\AA}, as  well as their errors, for each
Prism spectrum, and  the last row gives the  mean counts. These values
are uncorrected for Milky Way  extinction. There  is no evidence  that
the afterglow flux  varied during  the  course of these   observations
(about 5 hours). The  absence of flux recovery  shortward of the Lyman
edge at $z = 2.323$ indicates  a large \ion{H}{1}  column density.  We
return to this point in Sect.~\ref{sec:column} below.

The  spectra are  also  affected by  the  \textit{HST} and  STIS Point
Spread Function. Therefore we deconvolved the spectra using the method
described in Smette et al.\ (2001). The integrated fluxes in the range
3000 {\AA} $<\lambda<$ 3300 {\AA}  from the deconvolved spectrum along
with   the  estimated errors  are    given in  the   last  column   of
Table~\ref{tab:prism}.

\begin{table}[t]
\caption{Countrate and Integrated Fluxes from the four Prism Spectra}
  \label{tab:prism}
 \begin{center}
   \begin{tabular}{cr@{$\pm$}lr@{$\pm$}lr@{$\pm$}l}
\hline
\hline
&
\multicolumn{2}{c}{Countrate} &
\multicolumn{2}{c}{Flux} &
\multicolumn{2}{c}{Flux } \\
&
\multicolumn{2}{c}{photons/s} &
\multicolumn{2}{c}{} &
\multicolumn{2}{c}{(deconvolved)} \\
\hline
 & 0.245 & 0.017 & 2.80 & 0.32 & 3.81 & 0.40\\
 & 0.231 & 0.011 & 2.42 & 0.19 & 2.37 & 0.21\\
 & 0.235 & 0.011 & 2.48 & 0.20 & 2.85 & 0.24\\
 & 0.251 & 0.011 & 2.64 & 0.21 & 3.31 & 0.27\\
\hline
Mean &0.240 & 0.006 & 2.59 & 0.12 & 3.08 & 0.14\\
\hline
  \end{tabular}
 \end{center}
\tablecomments{The flux is given in units of
 $10^{-15}~\mathrm{erg}~\mathrm{cm}^{-2}~\mathrm{s}^{-1}$.}
\end{table}

\begin{table}[htb]
  \caption{Flux Densities Measured in the G430L Spectra}
  \label{tab:g430l}
 \begin{center}
 \begin{tabular}{cccc@{$\pm$}l}
\hline
\hline
    $\lambda_\mathrm{min}$ & $\lambda_\mathrm{max}$ & $\overline{\lambda}$ &
\multicolumn{2}{c}{Mean flux density} \\
({\AA}) & ({\AA}) & ({\AA}) &
\multicolumn{2}{c}{}\\
\hline
   3305.00 &  3355.00 &  3329.58 & 3.35 & 0.43\\
   3600.00 &  3650.00 &  3624.85 & 2.49 & 0.34\\
   3840.00 &  3890.00 &  3864.04 & 2.70 & 0.36\\
   4060.00 &  4110.00 &  4084.73 & 3.01 & 0.18\\
   4370.00 &  4420.00 &  4394.04 & 2.97 & 0.15\\
   4480.00 &  4530.00 &  4505.03 & 2.84 & 0.14\\
   4690.00 &  4740.00 &  4714.40 & 2.91 & 0.14\\
   4870.00 &  4920.00 &  4893.22 & 2.83 & 0.12\\
   4990.00 &  5040.00 &  5018.06 & 2.98 & 0.12\\
   5180.00 &  5230.00 &  5204.88 & 2.64 & 0.14\\
   5350.00 &  5400.00 &  5373.63 & 2.72 & 0.12\\
   5590.00 &  5640.00 &  5617.51 & 2.66 & 0.12\\
\hline
  \end{tabular}
 \end{center}
\tablecomments{Flux densities given in
 $10^{-17}~\mathrm{erg}~\mathrm{cm}^{-2}~\mathrm{s}^{-1}~\mathrm{\AA}^{-1}$,\\
 corrected from Milky Way extinction.}
\end{table}

\subsubsection{The G430L Spectra}
\label{sec:g430l_spectra}

The G430L  spectra cover the  spectral range  from 3000~\AA  \ to 5700
\AA. The purpose of the observation was to determine the optical slope
of the  afterglow spectrum with good  accuracy using near simultaneous
observations over 3000--16000 \AA \ together with  the NICMOS data. As
the GRB  redshift  was  larger than   foreseen when  the  Phase  2 was
submitted, the spectra  are   affected by Ly$\alpha$  absorption  (the
Ly$\alpha$ forest  starts at 4048    \AA) and hence  less useful   for
obtaining a precise measurement  of the  spectral slope. In  addition,
the spectra have a rather poor S/N and the  telescope was not dithered
between the 2 orbits as it was feared that the  afterglow could be too
faint  for the processing of  individual spectra. However, in order to
limit the effect  of the numerous  cosmic rays, 2 CR--SPLIT  exposures
were made in each of the 2 orbits (cf. Table~\ref{tab:journal}).

\begin{figure}[h]
 \begin{flushleft}
{\includegraphics[width=0.78\columnwidth,angle=90,clip]{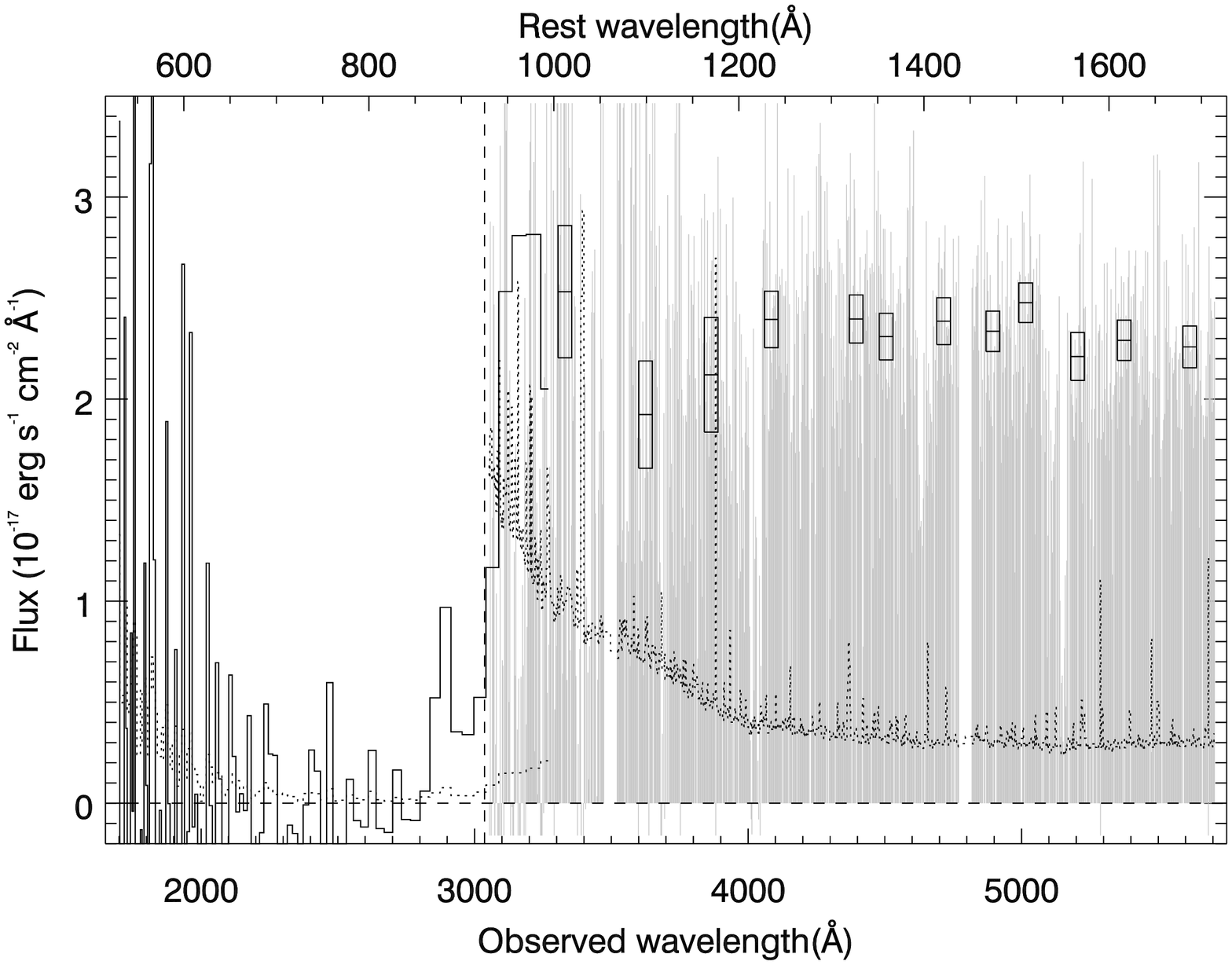}}
\caption{}{\textit{HST} STIS spectra of \objectname{GRB 021004} taken
on 2002 October 6. The deconvolved Prism spectrum and its 1$\sigma$ array is 
the histogram extending from below 2000 {\AA} to about 3300 {\AA}. The G430L 
spectrum is shown in gray extending from about 3000 {\AA} to 5500 {\AA}. The 
sharp drop in flux around 3000 \AA \ is the Lyman-limit absorption edge due 
to neutral hydrogen in the host galaxy. The boxes show the average flux 
densities in wavelength ranges free of strong Ly$\alpha$ forest lines. The
vertical extent of the boxes corresponds to the 1$\sigma$ error ranges.} 
 \label{fig:stis_spectra}
\end{flushleft}
\end{figure}

Due  to  the lack of  dithering, correct  processing  of  hot and warm
pixels is  crucial  but   in practice  very difficult.    The  reduced
individual  spectra still  show spikes.  They are  usually  one or two
pixel wide and often  common between the  different spectra.  A number
of criteria were defined to decide that a  given pixel in the combined
spectrum is not  valid  (i.e., due to wrongly   corrected warm or  hot
pixels) so that its value is not  used in subsequent analysis: (a) its
quality value is  larger or equal to  175,  which include  among other
conditions unrepairable hot  pixels,  pixels considered as  cosmics by
CR-SPLIT and saturated  pixels, and hot pixels (CCD  dark rate $>$ 0.2
counts/s); (b) pixels  with   'net' countrate  larger  than 0.09;  and
finally,  (c)   pixels   with 'net' countrate   smaller   than $-0.03$
(blemishes).   Figure  \ref{fig:stis_spectra}   shows   the  resulting
spectrum, its 1$\sigma$ error spectrum,  as well as boxes representing
the    mean values of   valid  pixels within  a    50 \AA \ range, and
corresponding standard error.   These wavelength ranges  were selected
away from the absorption lines  reported by Matheson  et al.\ (2003).  
Their limiting wavelengths as  well as their corresponding mean fluxes
and  errors,  corrected from  Milky  Way extinction,  are  reported in
Table~\ref{tab:g430l}. The best fitted power-law over these ranges has
a  slope of  $\beta  = 1.71\pm0.14$ with  a  reduced  $\chi^2 = 0.73$,
significantly steeper than  the $\beta \approx  1.0$ in the X-ray band
(see  below). In  Sect.~\ref{Sect:SED}  we analyze  the  full Spectral
Energy Distribution (SED) and conclude that the  steep slope in the UV
most likely   is due  to (modest) Small    Magellanic Cloud (SMC) like
extinction in the host galaxy. It is worth  noting that the integrated
flux of  the deconvolved Prism  spectrum  over 3000  {\AA} $<\lambda<$
3300   {\AA},  corrected      for      Milky  Way   extinction      is
$3.45\pm0.11\times10^{-15}~\mathrm{erg}~\mathrm{cm}^{-2}~\mathrm{s}^{-1}$,
while  the  corresponding    value in     the   G430L  spectrum     is
$3.15\pm0.46\times10^{-15}~\mathrm{erg}~\mathrm{cm}^{-2}~\mathrm{s}^{-1}$. 
The fact that these two values are consistent with each other gives us
confidence that the flux calibration of both spectra is correct.

\subsubsection{Limit on the \ion{H}{1} Column Density}
\label{sec:column}
Assuming that  the underlying   continuum  can be  represented by  the
extrapolation of the power-law described in the previous paragraph, we
measure  a  total equivalent width  of  $1.9\pm0.5$ \AA  \ using the 6
pixels covering the  range from 2417  to  2560 \AA \  (redwards of the
Lyman-limit  of the  foreground  $z=1.60$ \ion{Mg}{2}  absorber).  The
error  is only   the statistical error   and it  does not include  the
(likely larger)  systematic  error   from  the sky    subtraction  and
flat-fielding. For  a \ion{H}{1} column density below $1\times10^{18}$
cm$^{-2}$ at $z=2.33$, the continuum should  have recovered, giving an
equivalent width  above  4.5 \AA.  Assuming no  additional intervening
strong  (Lyman-limit) absorption, which  is reasonable as there are no
intervening  metal  line systems at   $1.60<z<2.33$  (M\o ller et al.\ 
2002), we conclude that the  \ion{H}{1} column density is larger  than
$1\times10^{18}$ cm$^{-2}$ (5$\sigma$). 

The G430L spectra also show the Ly$\alpha$ absorption lines reported
by (M\o ller  et  al.\ 2002), as   well as the associated  Ly$\beta$
lines.  However,  the latter appear in  a noisy part  of the spectrum,
with a S/N/pixel of about  2.5, which is  also affected by a number of
bad pixels.  M\o ller et al.\ (2002) find from the
Ly$\alpha$   absorption  line that the   \ion{H}{1}  column density is
constrained to be below $1.1\times10^{20}$~cm$^{-2}$.  When we use the
Voigt-profile parameters used to fit their  low resolution spectrum (a
total column density of $6\times10^{19}$ cm$^{-2}$ and a $b$ parameter
of  15 km  s$^{-1}$) we find   that the corresponding Ly$\beta$  lines
would appear weaker than the observed lines in the STIS spectrum.  The
significance of  this  finding  is  low  due to   the low S/N   of the
spectrum.    However, we can conclude  that,   either the total column
density is larger than $6\times10^{19}$ cm$^{-2}$ or the $b$ parameter
is larger 15 km s$^{-1}$.  
Fiore et al.\ (2004) find,
based on   a    high resolution  spectrum    obtained  with  the  UVES
spectrograph on the VLT, that the absorption profiles all contain very
wide components showing that the latter explanation  appears to be the
correct one.  From the UVES spectrum  the \ion{H}{1} column density is
constrained to be in the range 1--10$\times10^{19}$ cm$^{-2}$ 
(Castro-Tirado et al.\ 2005, in preparation).

\subsection{ACS and NICMOS Imaging Observations}
\label{sec:imaging}
The ACS observations from 2002 October 7 and 11 were CR-SPLIT observations
consisting of only one or two exposures. These were reduced through
the pipeline. The later epochs were observed with multiple exposures 
using non-integral multiple pixel dither steps and combined using 
MULTIDRIZZLE using pixfrac=1 and scale=0.66 (Fruchter \& Hook 2002; Koekemoer
2002).

All NICMOS images were taken using the NIC3 camera (51$\times$51 arcsec$^2$ 
field-of-view, 0.203 arcsec per pixel), MULTIACCUM mode (sample up the ramp), 
SPARS64 sample sequence and $\sim$3 arcsec dithers along each axis between 
exposures for bad pixel rejection. All data were taken in South Atlantic 
Anomaly free orbits except for the first two exposures of epoch 2.
These were taken in an orbit following the last shallow pass of the day 
and were only mildly impacted, so correction for cosmic-ray persistence was
not necessary. Two different calibration techniques were applied to this data. 
Epochs one, two and three (2002 October 7, 11 and 22, respectively) were 
calibrated with the standard NICMOS calibration pipeline CALNICA, using the 
best available reference files in the calibration database. Following CALNICA, 
the STSDAS task PEDSKY was used to remove any remaining quad-based DC biases 
("pedestal") signatures in the data. A sky image was constructed for each of 
the two filters using a source-masked median of all the epoch one and epoch 
two images combined, which were taken at different spacecraft orientations and 
dither positions. The sky image was then subtracted from each of the calibrated, 
PEDSKY corrected images.

For epochs four and five (2003 May 26 and July 21) a different technique was 
used, employing new methods developed during the Hubble Ultra Deep Field 
(HUDF) NICMOS analysis. First, a correction was made to the raw images to 
remove amplifier crosstalk from bright sources (also known as the "Mr. 
Stay-puft" anomaly). Although there are no bright sources in this field, every 
source produces a faint amplifier stripe and reflection to some level in the 
data. This is especially true of cosmic ray hits during the sample up the ramp 
MULTIACCUM exposures. Even though the cosmic ray hits themselves are flagged 
and rejected by the CALNICA processing, each hit leaves a faint stripe and 
reflection due to the amplifier effects, which can impact the noise floor. 
After this correction, the PEDSKY DC bias removal tool was run on dark-subtracted 
versions of each read up the ramp, and any measured bias offset (per quadrant) 
and sky background was subtracted from the raw reads. An image of the sky 
background from the HUDF was used as the sky model in pedsky rather than the usual
internal lamp flat. 
The entire MULTIACCUM sequence was then run through CALNICA as usual, 
except that the HUDF superdark was used instead of the standard dark reference 
file. This was possible because the same readout sample sequence that was 
used for the HUDF (SPARS64) was also used for these data.

\subsubsection{Photometry}
\label{sec:photometry}

Photometry was done using aperture photometry. In Table~\ref{tab:journal}
we provide measured countrates and the apertures sizes. For the 2002 
October 7 and 11
observations the countrates in Table~\ref{tab:journal} are corrected 
for geometrical distortion using the Pixel Area Map from the STScI 
web-pages\footnote{{\tt http://www.stsci.edu/hst/acs/analysis/zeropoints/analysis/PAMS}}.
For the subsequent epochs this correction is done automatically during 
drizzling. To convert the measured countrates to magnitudes we used the 
{\it synphot} package under STSDAS. {\it Synphot} calculates the AB magnitude 
corresponding to the measured countrate and the applied aperture size. AB 
magnitudes derived in this way are given in the last column of 
Table~\ref{tab:journal}. 

To accurately measure the afterglow component in the 2002 October 22 and 
2002 November 26 F606W observations, where the host galaxy is contributing 
a large fraction of the flux, we drizzled these images onto the 
coordinate system of the 2003 May 31 image. By subtracting the May 31 image 
we could then measure the afterglow component using a small aperture
with radius 0.2 arcsec\footnote{We note that another transient source, 
presumably a supernova, is detected in the May 31 image superimposed on 
a faint galaxy at the celestial position RA(2000) = 00:26:55.8, 
Dec(2000) = 18:55:28.}. 

For the NICMOS images we use a circular aperture with radius equal to 
5 pixels for the early points (October) and 2.5 pixels for the late points
(November and later) due to the larger pixels of the NIC3 detector (the pixel
scale is 0.203 arcsec per pixel). To derive AB magnitude we multiply
the counts with 1.075 (early) and 1.15 (late) to compensate for the finite 
aperture and use the most recent photometric keywords available at the 
\textit{HST} 
web-site\footnote{{\tt http://www.stsci.edu/hst/nicmos/performance/photometry/nic13\_postncs\_keywords.html}} to get the zero-points.

\subsubsection{The Late-time Lightcurve}
\label{sec:lightcurve}

The ACS photometry allows a precise measurement of the late-time decay slope of
the optical/near-IR afterglow lightcurve. There is a substantial disagreement 
between 
different reports in the literature. The values of the late-time decay slope 
ranges from $\alpha_2 = 1.43\pm0.03$ (H03 -- note, however, that they argue for 
$\alpha_2 = 1.98$ as the best value when including also the broadband SED in the  
analysis) to $\alpha_2 = 2.9$ (M03). H03 find a jet-break time of 
$4.74\pm0.14$ days, M03 find 9 days (no error bar) and Bj{\"o}rnsson
et al.\ (2004) 0.6 days. The main 
reasons for these large disagreements are the very ``bumpy'' 
lightcurve (making the determination of the epoch of the jet-break very
sensitive to the sampling of the lightcurve) and the relatively bright 
underlying host galaxy (potentially complicating the determination of the late 
time slope $\alpha_2$). M03 assume an overly bright host galaxy
magnitude (based on the November epoch of ACS observation in Table 1) and 
hence they derive an incorrectly large value of $\alpha_2$.
Rol et al.\ (2003) argue for a jet-break around 1 day based on the polarimetry,
whereas Lazzati et al.\ (2004) find that a later 
jet-break time of about 3$\pm$1 days is consistent with the polarimetry 
after day 1. Bj{\"o}rnsson et al.\ (2004) are able to fit both the optical
lightcurve, the radio and X-ray observations and the full evolution of the 
polarization using a model with four episodes of energy injection. In this
model the real jet break takes place around 0.6 days and the apparent later
break time is an artifact of the rebrightenings caused by energy injections.

\begin{figure}[h]
 \begin{flushleft}
{\includegraphics[width=0.78\columnwidth,angle=90,clip]{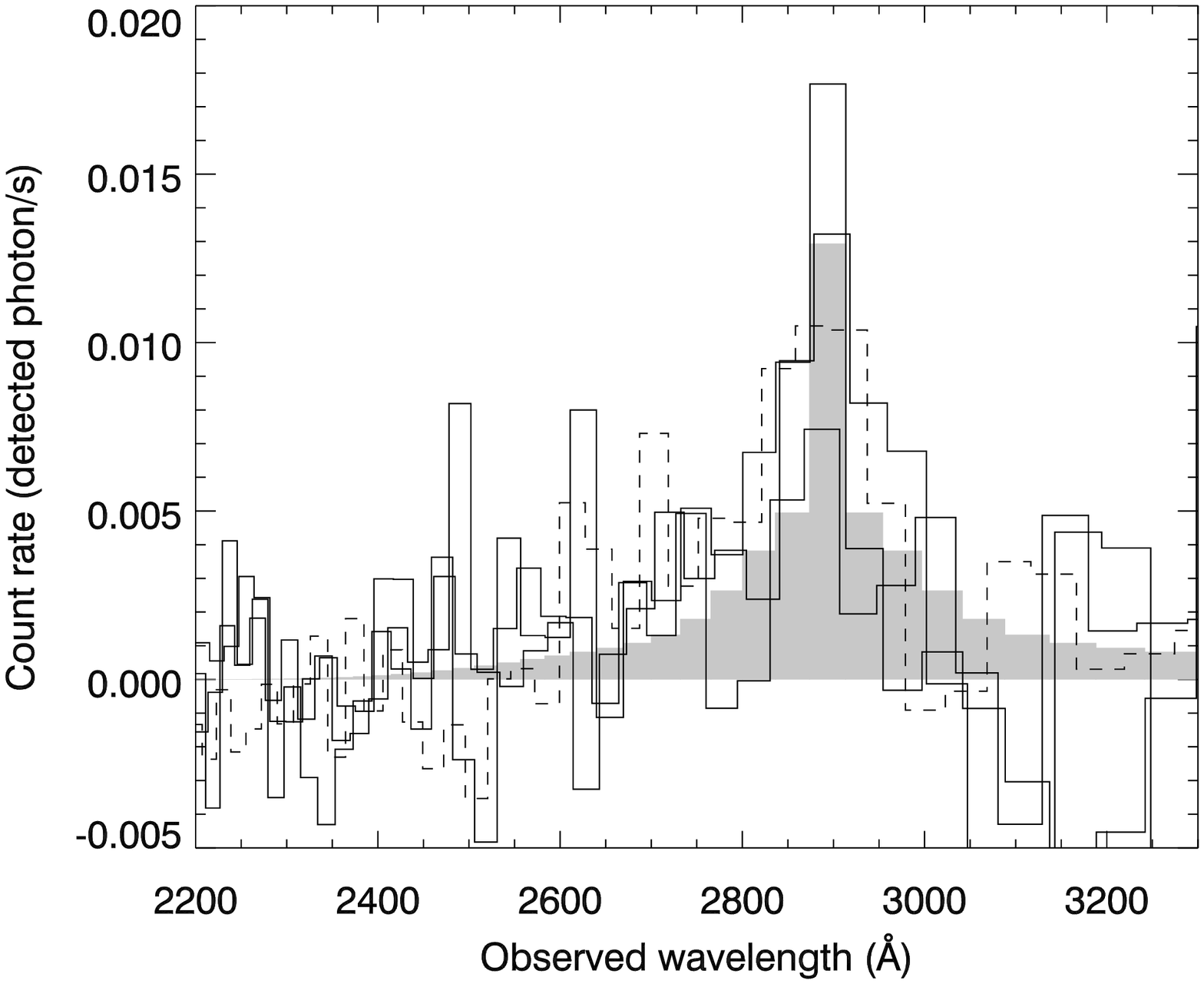}}
\caption{}{Residual of the subtraction of the individual Prism spectra
by a model of the afterglow spectrum with a Lyman break convolved by the
\textit{HST+STIS} LSF. The excess emission is clearly seen in all four
individual spectra (grey and black full drawn and dashed curves) and appear
very similar in shape to a unresolved emission line convolved by the
\textit{HST+STIS} LSF (shown as filled grey histogram). This line may be due a
$z = 1.38$ Ly$\alpha$ emission line with a flux of $6\times
10^{-16}~\mbox{ergs} ~ \mbox{s}^{-1}~\mbox{cm}^{-2}$.  }
 \label{fig:excess}
\end{flushleft}
\end{figure}

\begin{figure}[h]
 \begin{flushleft}
{\includegraphics[width=1.00\columnwidth,angle=0,clip]{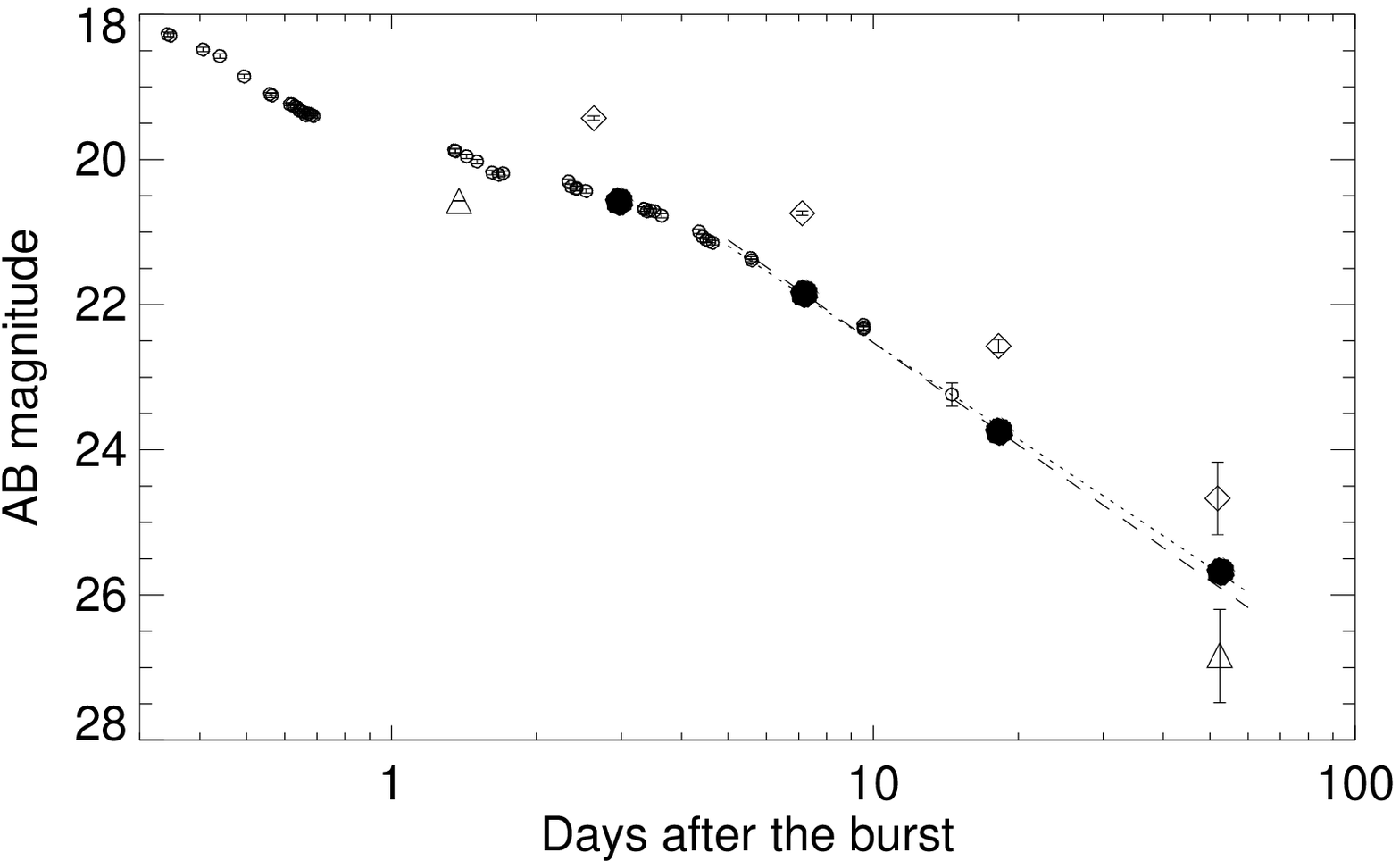}}
\caption{}{
The lightcurve from 0.3 days to 53 days after the GRB. The \textit{HST} 
points are shown
with filled circles (error bars are smaller than the symbol sizes). The 
small, open circles are the ground based R-band data from H03
shifted to F606W AB magnitudes by adding 0.25 mag and corrected 
for a host magnitude of R(AB) = 24.40. The lightcurve seems to be 
flattening at late times, which could be due to the transition to 
non-relativistic expansion, but also due to bumps in the lightcurve
around 2002 October 11. We also plot the four F160W points and the two
X-ray points (arbitrarily shifted in flux) as diamonds and triangles 
with error-bars. These points indicate an achromatic evolution of
the afterglow from the H-band to the X-ray band over a period of up to 
50 days after the GRB.
}
 \label{fig:lightcurve}
\end{flushleft}
\end{figure}

The epoch 5 observation from May 31 2003 is more
than half a year after the explosion, and the afterglow light at this epoch
is most likely negligible. Conservatively assuming a late-time decay-slope 
of $\alpha_2 = 1.7$ we infer an afterglow magnitude of 28.5 in 
the F606W band at 2003 May 31. The decay-slope $\alpha_2$ is larger than 1.7
(see below) and the afterglow magnitude at 2003 May 31 will therefore be 
fainter than 28.5. The rightmost column in Table~1 contains the 
afterglow magnitudes based on the assumption of no afterglow
emission in the last epoch images. In Fig.~\ref{fig:lightcurve} we plot the
F606W magnitudes for epochs 1--4. Clearly, the four points are not 
consistent with a single power-law. Epoch 1 is before the 
break time as determined both by H03 and M03 
and we can hence exclude this point from the power-law fit. However, 
a single power-law fit to the last three points is also formally rejected 
($\alpha_2 = 1.86$, $\chi^2 = 9$, 1 d.o.f). The dotted line shows the result 
of an unweighted fit to the epoch 2--4 points, and the dashed line shows the 
line defined by the epoch 2 and 3 points. The dotted and dashed lines 
correspond to $\alpha_2 = 1.77$ and $\alpha_2 = 1.88$, respectively. 

The last \textit{HST} detection indicates a flattening of the late-time 
afterglow. This effect cannot be caused by residual afterglow emission in
the epoch 5 image as this would have the opposite effect, making the
lightcurve bend the other way. The flattening could be the result
of the early transition to non-relativistic expansion. According to 
Livio \& Waxman (2000) $\alpha_2$ should evolve towards $\sim$0.9 on a 
timescale of 5 months. An underlying
supernova (Stanek et al.\ 2003; Hjorth et al.\ 2003a) is not expected to 
contribute significantly to the restframe UV light corresponding to the 
ACS filters. An alternative explanation is a bump in the light curve 
around 2002 October 11 causing the afterglow to be too
bright. Given the very bumpy nature of the GRB~021004 lightcurve this
does not appear unlikely. 

\subsubsection{The NIR/Optical SED of the Afterglow}
\label{Sect:SED}

\begin{figure}[h]
 \begin{flushleft}
{\includegraphics[width=0.50\columnwidth,angle=0,clip]{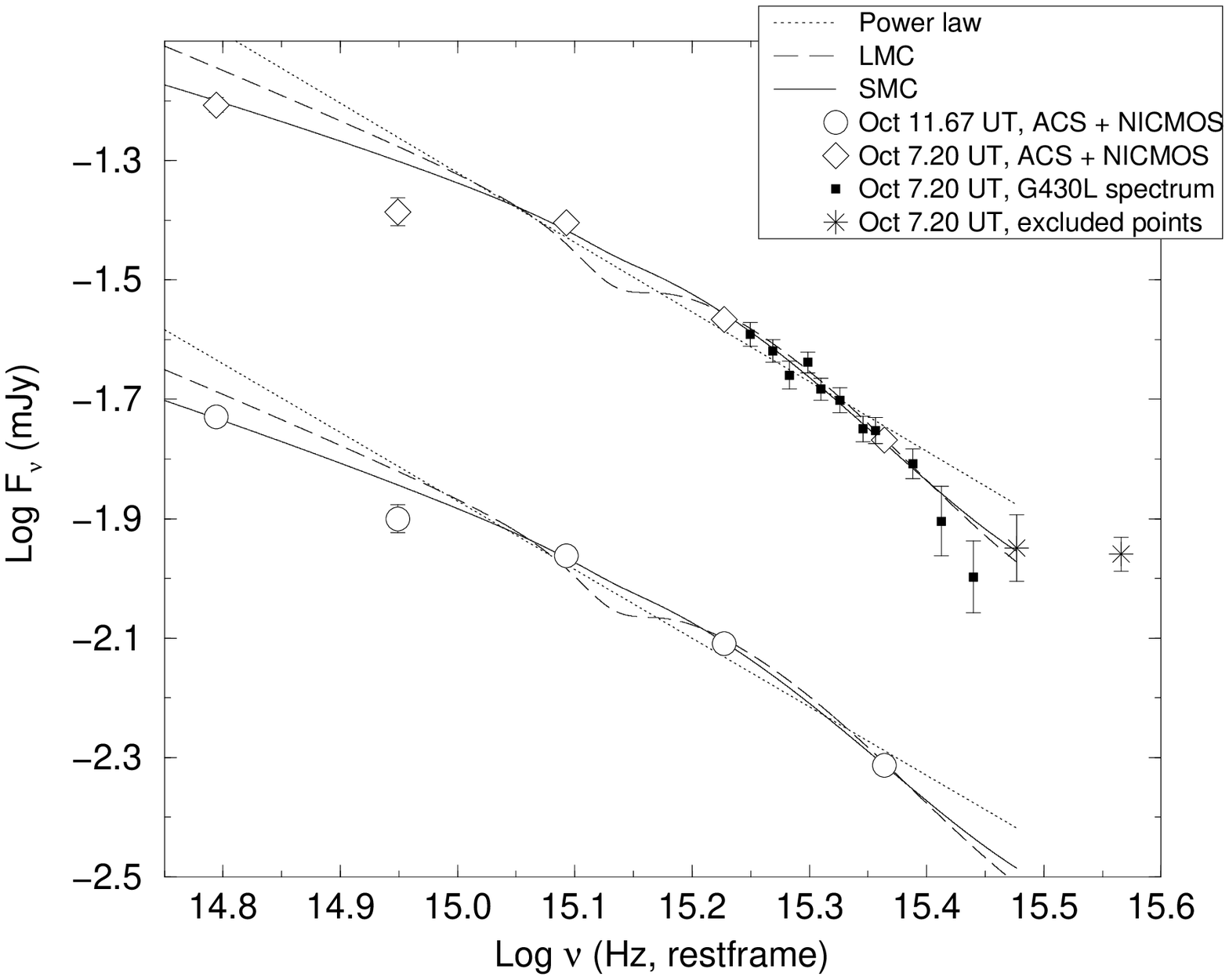}}
{\includegraphics[width=0.50\columnwidth,angle=0,clip]{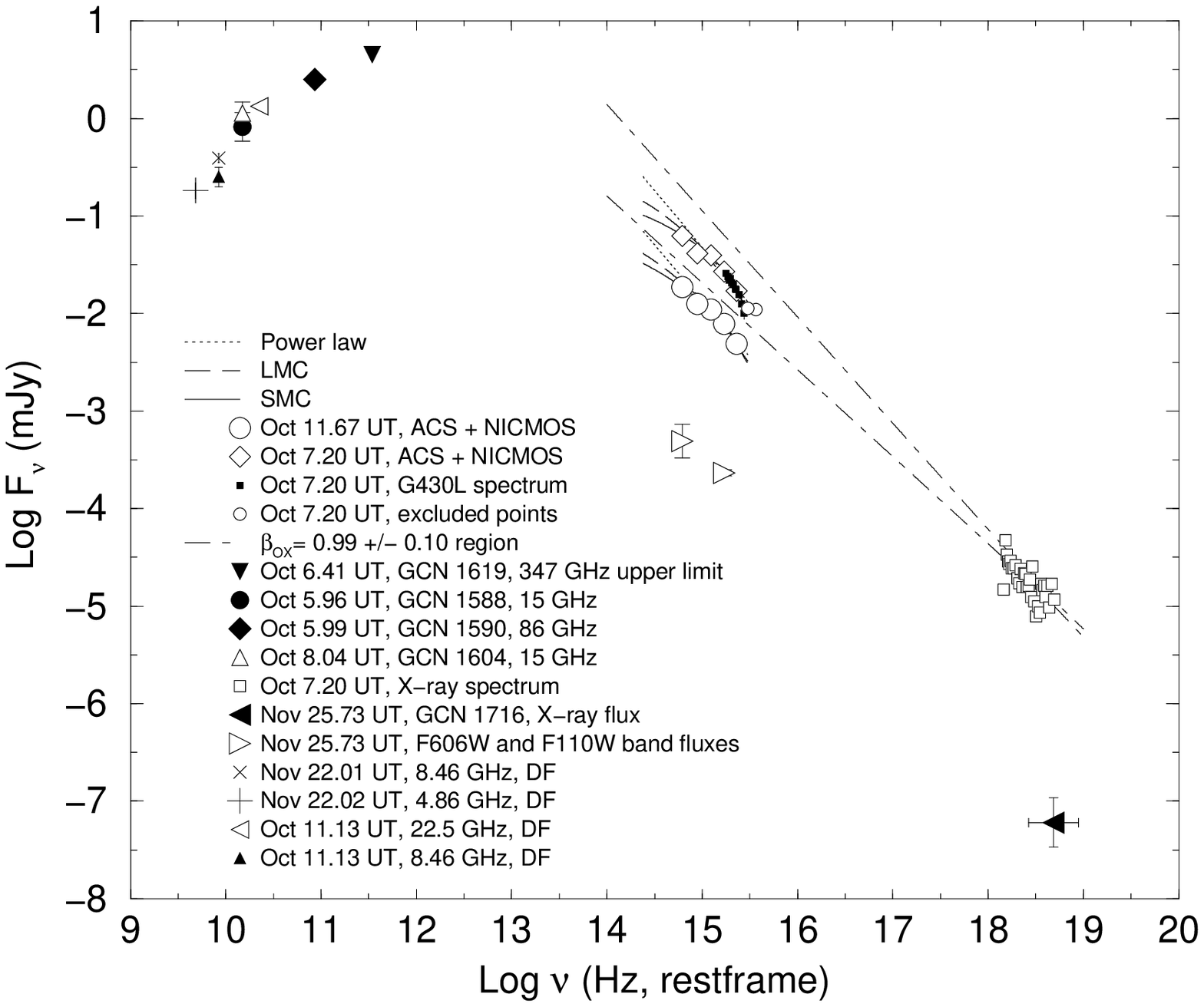}}
\caption{}{{\em  Upper panel:} The  SED  of the  afterglow  on 2002 October 7.2
(diamonds) and 2002 October 11.67 (circles).The fluxes derived from the bins of
the STIS G430L spectrum  (see Table~3) have been shifted to 2002 October 7.2
and are shown with filled squares. The stars show the two data points (the
F250W and the bluest bin of the  STIS G430L spectrum) with a restframe
wavelength  below 1000 \AA. These are not been included in the fits.  The  fit
obtained with a MW-like  extinction law is not displayed  since it provides
unphysical negative  $A_{\rm V}$ values (see Table~\ref{Table:SED}).  {\em
Lower panel}: The broadband SED of the GRB 021004 afterglow, from radio to
X-rays.  The plot represents the SED at Oct 2002 7.2, 11.67 and 2002 November
25.73. The four last radio data points,  around Oct 11 and Nov 22, where
collected from the compilation by Dale Frail\footnote{{\tt
http://www.aoc.nrao.edu/$\sim$dfrail/allgrb\_table.shtml}} and are indicated
with ``DF''. The broad band  SED from  the optical  to the  X-ray range  has
roughly  a constant spectral   index  ($\beta_\mathrm{OX} \approx 1$). The
fact that the optical to X-ray slope $\beta_\mathrm{OX}$ is the same within the
errors from October 7 and until November 26 with a value close to $\beta_X$
means that the cooling frequency must have remained constant and close to the
optical in this time span.} The dot-dashed line represents the 1$\sigma$ region
of the extrapolated $\beta_\mathrm{OX}=0.99 \pm 0.1$ power-law fit derived by
H03.  
\label{fig:SED} 
\end{flushleft} 
\end{figure}

As shown in Fig.~\ref{fig:SED}, the \textit{HST} observations allow us to 
construct the SED of the afterglow around 2002 October $\sim$7.20 and 
$\sim$11.67. Given that the afterglow photometry is based on images where 
the host galaxy has been subtracted, the SED is not affected by host 
galaxy light contamination.

The \textit{HST} photometry was shifted to a common epoch at 2002 October 7.20 
and 11.67 assuming power-law decay slopes of $\alpha = 0.85$ (from H03)
and $\alpha = 1.86$ (based on the analysis described above), 
respectively. The two observation sets are well clustered
around these two common dates (maximum epoch shifts of $\delta t= $ 0.28
days for October $\sim$7.20), so the derived SEDs are not very sensitive
to the assumed values of $\alpha$ before the lightcurve break at $\sim$7.20 
(e.g., $\delta m <0.16$ for $0.5 < \alpha < 2.0$).
We have added in quadrature a 0.05 mag error to the F110W-band magnitudes
displayed in Table~\ref{tab:journal} in order to account for the NIC3
intra-pixel sensitivity variations. These sensitivity variations are 
especially relevant in the F110W-band and can produce sensitivity variations as 
large as $30\%$ (peak-to-peak, Storrs et al.\ 1999).

We  analyze  the SED  by  fitting  a function  of  the  form $F_{\nu}  \sim
\nu^{-\beta} \times 10^{-0.4 A_{\nu}}$, where $\beta$ is the spectral index
and  $A_{\nu}$ is  the extinction  at frequency  $\nu$. $A_{\nu}$  has been
parametrized in terms  of $A_{\rm V}$ following the  three extinction laws
Milky Way (MW), Large Magellanic Cloud (LMC), and SMC given by Pei et al.\ 
(1992). In the analysis we have included the UV
flux from the STIS G430L spectrum (Table~\ref{tab:g430l}). The Pei et  
al.\ (1992) extinction laws  are very uncertain for restframe wavelengths  
below 1000 \AA, so the bluest bins of the STIS G430L spectrum were not 
included in the analysis. For the same reason, and also due to the 
Ly$\alpha$ blanketing and Lyman-limit absorption, the F250W data point 
was also not included in the SED fit. As a consistency check we have 
corrected the F250W data point for Ly$\alpha$ blanketing by convolving 
the F250W filter sensitivity curve with a Ly$\alpha$ blanketing model and 
a Lyman-limit break corresponding to the burst redshift. The two excluded 
data points are roughly compatible with the optical/NIR SED (the two stars in
the upper panel of Fig.~\ref{fig:SED}).

For both epochs  the SEDs are clearly curved  and hence highly inconsistent
with a  pure power-law  spectrum ($\chi^2/\mathrm{d.o.f.}  \geq  16.7$, see
Table~\ref{Table:SED}). The LMC, and especially the MW extinction law which
yields $A_{\rm V} <0$, provide unacceptable fits.  Only the SMC extinction
law provides moderate $\chi^2/\mathrm{d.o.f.}$  values. The $A_{\rm V}$ and
$\beta$ values  derived for the SMC  ($\beta =0.30 \pm 0.06$,  $A_{\rm V} =
0.23 \pm 0.02$ at October 7.20 and $\beta =0.42 \pm 0.06$, $A_{\rm V} = 0.20
\pm 0.02$ at October 11.67) are consistent with the ones reported by H03
at 2002 October 10.072:  $\beta =0.39 \pm 0.12$, $A_{\rm V} = 0.26
\pm 0.04$.

The evolution of the afterglow is achromatic within the errors 
from 4000 \AA \ to 16000 \AA \ over the period from October 7 to
October 22 (3 to 18 days after the burst). The F606W$-$F160W color from 
November 25--26 (53 days after the burst) observation is
consistent with the color from the earlier epochs, but the error-bar
is very large. H03 find that R$-$I is constant from
0.35 days to 5.5 days after the burst. On the other hand, Matheson et al.\
(2003) and Bersier et al.\ (2003) present evidence for color changes 
in the near-UV and optical range during the first four nights. The
only way to reconcile these observations with ours is that the color 
changes must be related to a stochastic phenomenon superimposed on 
the normal
afterglow light whose color remains very stable. Moreover, the 
cooling break cannot have passed through the optical range during 
the period from 0.35 to 18 days after the burst and  
possibly not earlier than 53 days.

\begin{table}[t]
\caption{Fits to the NIR/Optical SED}
\label{Table:SED}
\begin{center}
\begin{tabular}{ccccc}
\hline
\hline
 Extinction  & Epoch UT & $A_{\rm V}$  & $\beta$ &$\chi^2/\mathrm{d.o.f.}$\\
 Law     & 2002 Oct.\ &        &     &       \\
\hline
 Unextincted & $7.20$ & $    0  $  & $1.17\pm0.02$ & $18$\\
 MW      & $7.20$ & $<0$ & $1.30\pm0.03$ & $19$\\
 LMC     & $7.20$ & $ 0.32\pm0.03$ & $0.44\pm0.06$ & $12$\\
 SMC     & $7.20$ & $ 0.23\pm0.02$ & $0.30\pm0.06$ & $3$\\
\hline
 Unextincted & $11.67$ & $   0   $ & $1.15\pm0.02$ & $53$\\
 MW      & $11.67$ & $<0$ & $1.26\pm0.03$ & $74$\\
 LMC     & $11.67$ & $ 0.30\pm0.03$ & $0.47\pm0.06$ & $31$\\
 SMC     & $11.67$ & $ 0.20\pm0.02$ & $0.42\pm0.06$ & $4$\\
\hline
  \end{tabular}
 \end{center}
\tablecomments{The table shows the SED fits for two epochs at October 7.20 and
 Oct 11.67.  For each epoch the three extinction laws (MW, LMC and SMC)
 given  by  Pei et  al.\  (1992)  were  considered.  For  completeness  the
 unextincted pure  power-law  is also included.  For both epochs,  the best
 solution is clearly obtained with a SMC extinction law.}
\end{table}

\subsubsection{Broadband SED} The X-ray afterglow of GRB~021004  was observed
with the Chandra X-ray Observatory on 2002 October 5.9 and 2002 November 25
(Sako \&  Harrison 2002a,b)  and is analyzed  in H03  and Fox  et al.\  (2003).
The  X-ray spectra  show no  evidence for  absorption in  addition  to the
foreground Galactic absorption  with a  90\% confidence upper  limit on  the
absorbing column  density  in   the  host  galaxy  or  along   the
line-of-sight  of 2.3$\times$10$^{21}$ cm$^{-2}$. H03 report a spectral slope
in the X-rays of  $\beta_X = 1.06\pm0.06$ and $\beta_X = 0.94\pm0.03$ fitting
to  2--10 keV  and 0.4--10 keV  respectively. On November  25 (52.3 days
after   the    burst)    the   X-ray    flux    was   reduced    to
$7.2\pm2.5\times10^{-16}$    erg    cm$^{-2}$    s$^{-1}$    compared    to
$4.3\times10^{-13}$ erg cm$^{-2}$ s$^{-1}$ at  around 1.5 days (Fox et al.\
2003).  This  corresponds  to a  temporal  decay  slope  in the  X-rays  of
$\alpha_X = 1.80\pm0.10$ -- very similar  to what we find for the late-time
slope in the  optical/near-IR. This means that the broad-band SED from the
optical to the X-ray band has a roughly constant shape $\sim$2 days and
$\sim$52  days after  the burst.  In Sect.~\ref{sec:discuss}  we return  to the
interpretation of the  broad-band SED within the context  of the blast wave
(fireball) and cannonball models.

In the lower panel of Fig.~\ref{fig:SED} the broadband SED from radio to 
X-rays is plotted. The epochs cluster around 2002 October 10.7,
11.67, and 2002 November 25.73. All the  radio measurements and  the X-ray
flux of November 25.73 are taken from the literature.  The X-ray
points of October 7.20 (empty squares in the lower panel of Fig.~\ref{fig:SED}) 
are based on our independent analysis of the Chandra X-ray spectrum taken
on Oct 5.4--6.4 and shifted in time to October 7.20  assuming a decay of
$\alpha=1.86$ as derived from the optical light curve.

\subsubsection{The Host Galaxy}

\begin{figure*}
{\includegraphics[width=1.00\columnwidth,angle=0,clip]{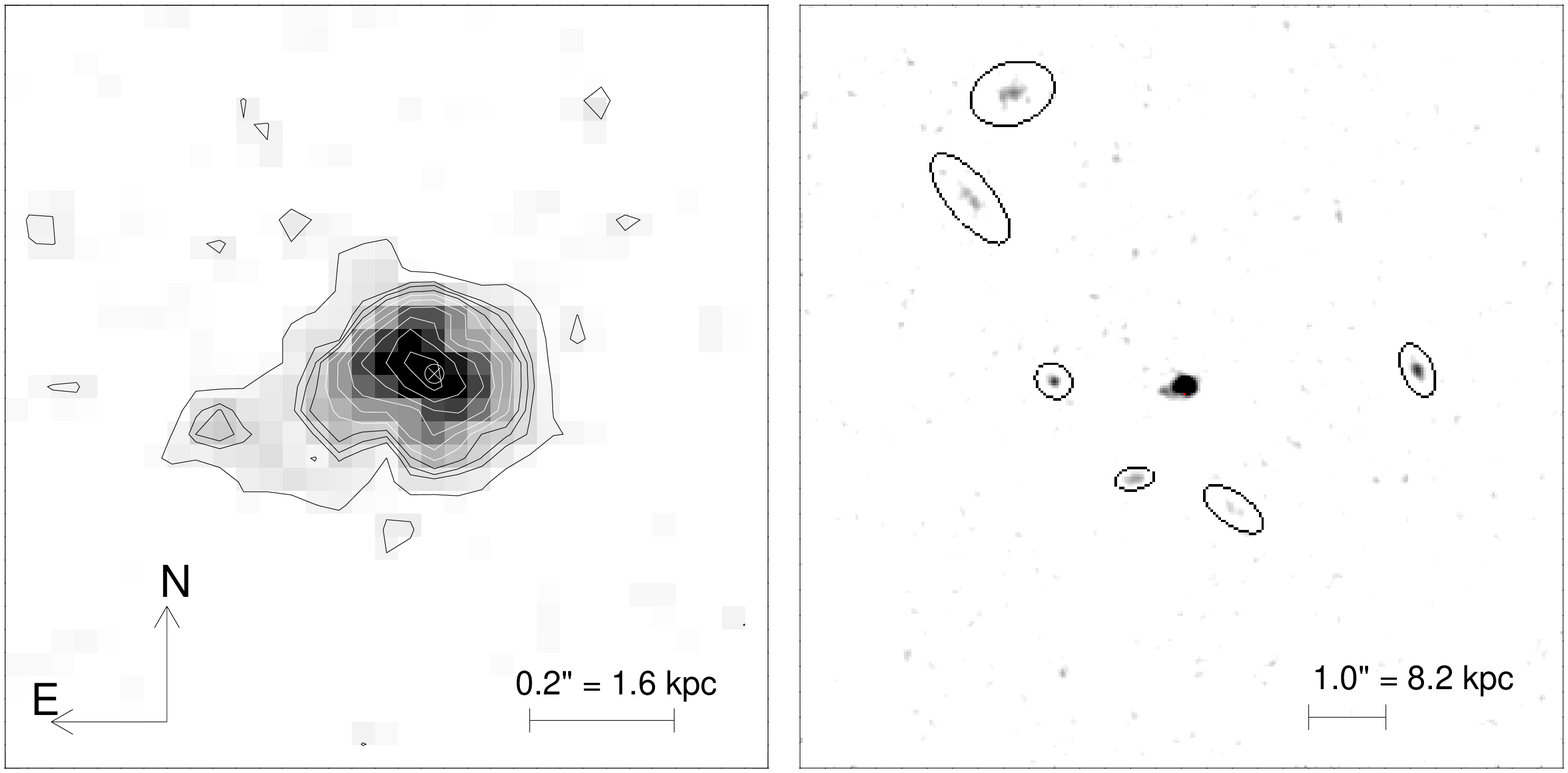}}
\caption{}{
{\it Left panel:} A 1$\times$1 arcsec$^2$ section of the ACS/F606W image from 
2003 May 31 around the host galaxy of GRB~021004. 
The GRB went off near the center of the galaxy (position
marked with a cross and an error circle). To better show the morphology
of the object we have overplotted contours with a logarithmic scaling.
{\it Right panel:} A smoothed version of the ACS/F606W image from
003 May 31 covering 10$\times$10 arcsec$^2$ centered on the host galaxy.
Some of the six very faint ($ \gtrsim 26$ mag, marked with ellipses) 
galaxies within 5 arcsec from the line-of-sight to the host galaxy
could be counterparts of the foreground absorbers. 
}
 \label{fig:host}
\end{figure*}

The host galaxy is clearly detected in all bands in the latest epoch images
from 2003 May--July. In  the left panel  of Fig.~\ref{fig:host} we  show a
1$\times$1 arcsec$^2$  section of the F606W  image from 2003 May 31 around
the host galaxy. The host galaxy has  a very compact core with a half light
radius of only  0.12 arcsec. The resolution of the ACS  image is about 0.05
arcsec so  the core is well resolved. The galaxy  has a faint
second  component  offset  by  about  0.28 arcsec  (2.2  kpc)  towards  the
east. We cannot, based on the imaging alone, determine if this second 
component is part of the host or due to an independent object, e.g. one of the 
foreground absorbers. 

The position  of the afterglow is  marked with a cross and the 5$\sigma$
error is  shown with  a circle  (1$\sigma$ =  0.08 pixels).  The afterglow
position  is offset  from  the galaxy  core  by only  0.4 drizzled  pixels,
corresponding to 0.015 arcsec or 119 parsec at $z=2.33$. This is one of the
smallest impact parameters measured so far (Bloom, Kulkarni \& Djorgovski 
2002), and although it could be a chance projection it suggests that 
the progenitor could be associated with a circumnuclear starburst. 

The right panel of Fig.~\ref{fig:host} shows a larger portion, 
10$\times$10 arcsec$^2$, of the field around the host galaxy. A 
number of very faint (R $>$ 26) galaxies are seen. Some of these are likely
associated with the strong intervening \ion{Mg}{2} absorbers at redshifts
$z=1.38$ and $z=1.60$ seen in the afterglow spectrum (M\o ller et al.\
2002; Castro-Tirado et al.\ 2004; see also Vreeswijk, M\o ller \& Fynbo 
2003; Jakobsson et al.\ 2004).

\begin{figure}
 \begin{flushleft}
{\includegraphics[width=1.00\columnwidth,angle=0,clip]{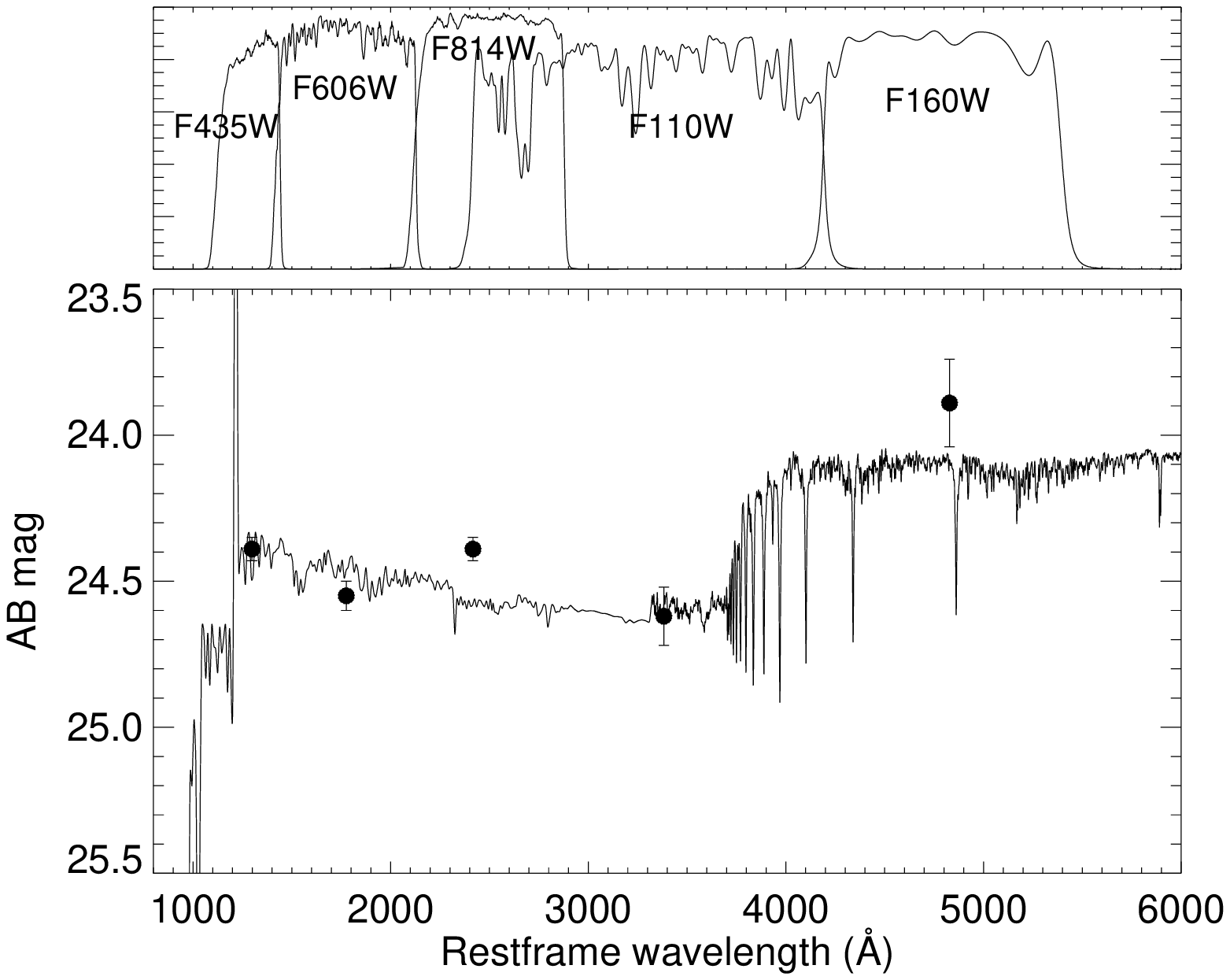}}
\caption{}{
The SED of the host galaxy extending from
about 1200 \AA \ to 5000 \AA \ in the restframe. Also shown is the
spectral synthesis model of a 42 Myr old star burst with $Z=0.004$
and with a Ly$\alpha$ emission line added in by hand. The top
panel shows the filter transmission curves corresponding to the
five photometric points.
}
 \label{fig:hostphot}
\end{flushleft}
\end{figure}

The  SED of  the galaxy  can be  quite well
constrained from  restframe 1200 \AA  \ to  5000 \AA \ based  on the  ACS and
NICMOS detections. This  range brackets the UV region  sensitive to young O
and  B stars, and  hence the  star formation  rate, and  the Balmer jump,
sensitive to the  age of the starburst. In  Fig.~\ref{fig:hostphot} we show
the  SED  of the host. The  SED is  essentially  flat  bluewards of  the
Balmer jump, and the Balmer jump has a magnitude of about 0.4 mag. Also
shown is the simulated spectrum of a starburst from the 2003 version of the
spectral synthesis models of Bruzual \& Charlot (2003). The model starburst
has an  age of 42 Myr  and a metallicity of  20\% solar. It  is possible to
obtain  reasonable fits with  ages between  30 and 100 Myr dependent  on the
metallicity. In addition the range  of allowed models could be increased by
allowing for extinction, but as the galaxy is known to have high equivalent
width Ly$\alpha$  line emission (M\o ller  et al.\ 2002)  there is little
room for extinction (e.g., Charlot \& Fall 1993). Based on the strength
of the UV continuum of the host galaxy we can derive a value of the 
Ly$\alpha$ equivalent width (M\o ller et al.\ 2002 give a lower limit).
The magnitude of the host galaxy of F606W$_\mathrm{AB}$ = $24.39\pm0.04$ 
corresponds 
to a specific flux of $F_{\lambda} = 1.1\times10^{-18}$ erg s$^{-1}$ 
cm$^{-2}$ \AA$^{-1}$. With the measured line-flux of $F_\mathrm{Ly\alpha} =
2.5\pm0.5 \times 10^{-16}$ erg s$^{-1}$ cm$^{-2}$ we derive an observed
equivalent width of $231\pm47$ \AA \ corresponding to $69\pm14$ \AA \ in
the restframe, very similar to what is found for the host galaxy of
GRB~000926 at $z=2.04$ (Fynbo et al.\ 2002). 

\subsubsection{Star Formation Rate}

The specific luminosity of the UV continuum provide a measurement of the
SFR. We use the relation of Kennicutt (1998)

\[
SFR(\mathrm{M_{\sun}\: yr^{-1}}) = 1.4\times10^{-28} L_{\nu}, 
\]
\noindent
where $L_{\nu}$ is the specific luminosity in units (erg s$^{-1}$ Hz$^{-1}$) 
in the 1500--2800 \AA \ range.
The observed AB magnitude of 24.55 corresponds to
$F_{\nu} = 5.50\times10^{-30}\; \mathrm{erg\; s^{-1}\; Hz^{-1}\; cm^{-2}}$. 
In our choice of cosmology this corresponds to a value of 
$L_{\nu} = 4 \pi d_l^2 F_{\nu}/(1+z) = 6.93\times10^{28}\; \mathrm{erg\; s^{-1}\; 
Hz^{-1}}$. This gives a SFR of $10\; \mathrm{M_{\sun}\; yr^{-1}}$. Due to dust
absorption this is a lower limit to the SFR, but as mentioned above
there is only little room for dust. We can also derive an estimate
for the SFR based on the Ly$\alpha$ flux. M\o ller et al.\ (2002)
report a Ly$\alpha$ flux of $2.46\pm0.50\times10^{-16}\; \mathrm{erg\; s^{-1}\;
cm^{-2}}$. This corresponds to a luminosity of $1.0\pm0.2\times10^{43}\;
\mathrm{erg\; s^{-1}}$ and a SFR of $10\; \mathrm{M_{\sun}\; yr^{-1}}$ 
(following the calculation in Fynbo et al.\ 2002). A similar value was
found by Djorgovski et al.\ (2002). The values of the SFR 
measured from the UV continuum and from the Ly$\alpha$ flux are
hence in good agreement as for GRB~030323 (Vreeswijk et al.\ 2004). Hence, 
at least for some high-$z$ starbursts Ly$\alpha$ emission is as reliable 
as the UV-continuum as SFR estimator (see Mas-Hesse et al.\ 2003 and 
Kunth et al.\ 2003 for critical discussions of the use of Ly$\alpha$ as 
SFR indicator).

\begin{figure}[h]
 \begin{flushleft}
{\includegraphics[width=0.80\columnwidth,angle=0,clip]{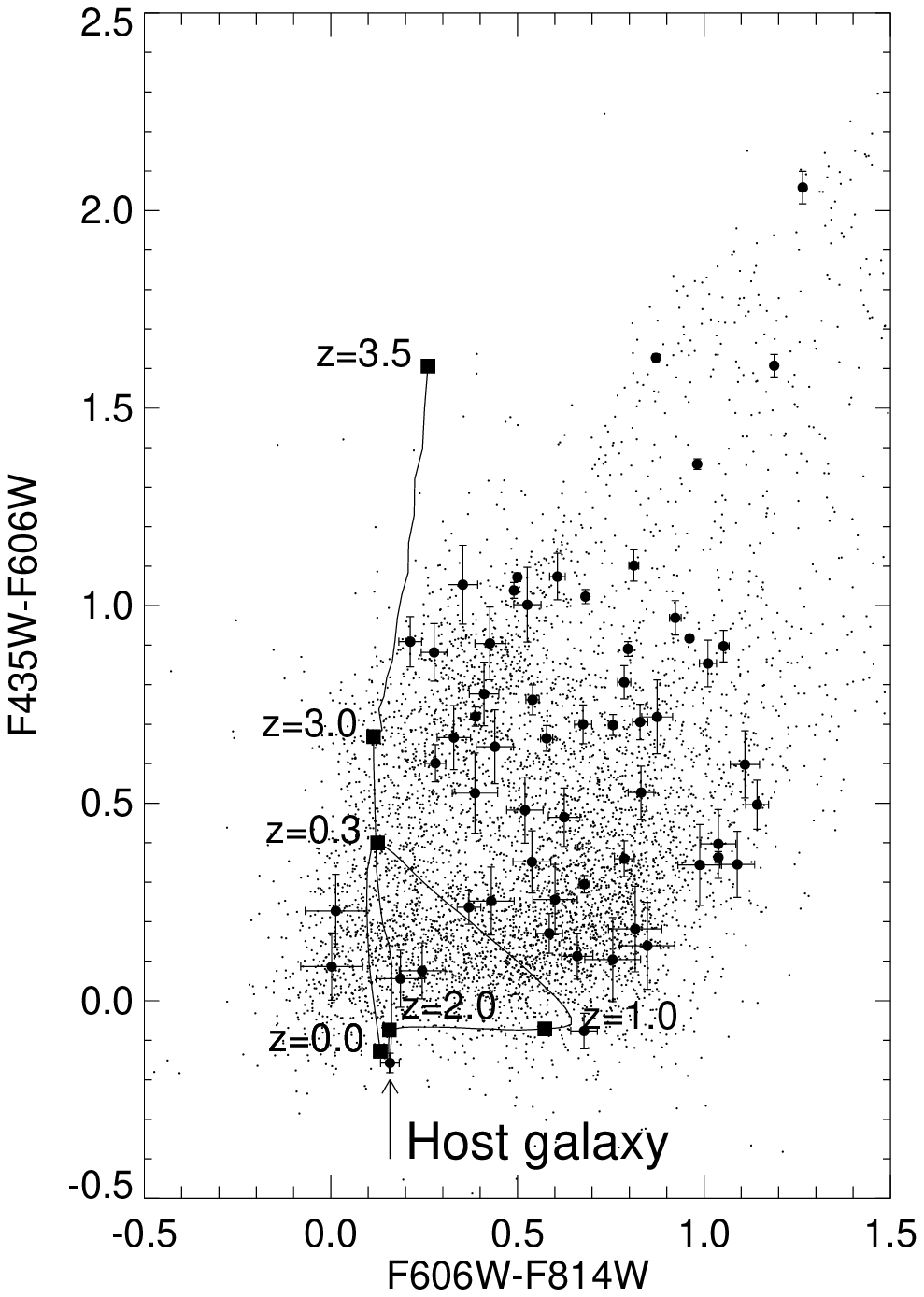}}
\caption{}{
The color of the host compared to the colors of galaxies in its environment 
(filled circles with error bars) and to galaxies in the GOODS South field 
(dots). Also shown are the colors of the spectral synthesis model shown in 
Fig.~\ref{fig:hostphot} as a function of redshift from $z=0$ to $z=3.5$ 
renormalized to go through the host galaxy point.}
\label{fig:hostenviron}
\end{flushleft}
\end{figure}

\section{Discussion and Conclusions}
\label{sec:discuss}

\subsection{The \ion{H}{1} Column Density}
The STIS spectroscopy is consistent with and complements a the measurements
from ground based near-UV and optical spectroscopy. We detect the 
Lyman-limit break associated with the GRB~021004 absorption system also 
detected in 
the Ly$\alpha$ resonance line and several metal lines at observer-frame 
optical wavelengths. Based on the Lyman-limit break we have placed a lower 
limit of about $1\times10^{18}$~cm$^{-2}$, significantly higher than the 
$\sim2\times$10$^{16}$ cm$^{-2}$ inferred by M03 based
on the Lyman series. The low column densities derived by M03
are caused by their assumption that the lines are not saturated.
From the analysis of the  X-ray afterglow Fox et al.\ (2003) derive
an 90\% confidence upper limit on the absorbing column density in the host 
galaxy or along the line-of-sight of 2.32$\times$10$^{21}$ cm$^{-2}$. 
The \ion{H}{1} column density is quite low
compared to that found for other GRB host galaxies (Vreeswijk et al.\
2003, their Fig.~4 and discussion thereof). Most of the Ly$\alpha$ 
lines from GRB afterglows are strongly damped and have inferred column densities
well above 10$^{21}$ cm$^{-2}$. Only GRB~021004, GRB~011211 (Vreeswijk et
al. 2005) and possibly GRB~030226 (Klose et al.\ 2004) depart from
this rule having column densities below the classical limit of 
2$\times$10$^{20}$ cm$^{-2}$ \ion{H}{1} for Damped Ly$\alpha$ Absorbers (DLAs)
in QSO spectra.  The Ly$\alpha$ absorption profile is 
furthermore peculiar by having multiple components
spread over about 3000 km s$^{-1}$ with clear evidence for line-locking between 
the components (Savaglio et al.\ 2002; M\o ller et al.\ 2002; Schaefer et al.\
2003; M03; Castro-Tirado
et al., in preparation). The line-locking implies a very strong radiation 
field along the line of sight, presumably from the GRB progenitor star/star
cluster, and it is possible that this radiation field 
has ionized the \ion{H}{1} along the line thereby explaining the low total 
column density in the host galaxy.
An alternative explanation for the low 
foreground column density compared to previously studied GRB lines-of-sight 
could be that the progenitor was located in the perimeter of the host galaxy.
However, the position of the afterglow very close to the host (although in
projection) in Fig.~\ref{fig:host} does not support this explanation.

\subsection{A Ly-$\alpha$ emission line at $ z = 1.38$?}
\label{sec:lya_138}

We  noted in Sect.~\ref{sec:prism_spectra} that the PRISM   spectrum seems to
show  an excess of emission  shortward  of the  Lyman-break.  As  the different
methods to derive the HI  column density at the  GRB redshift are consistent
with each other, it is unlikely that we  overestimate its value. Therefore, a
low HI column density cannot explain the shape of the Lyman-break.

One possibility is that this excess emission could be caused by  a Ly$\alpha$
emission line at $z = 1.38$, probably associated with absorption line system A
following the naming convention in M\o ller et  al.\ (2002).  In order to 
examine if this explanation is plausible, we have subtracted a model  of the
afterglow spectrum based on the results in Sect.~\ref{Sect:SED} from the
observed spectra.  In addition, \ion{H}{1}  continuous  absorption  has   been
added   to   model the Lyman-break, with an  \ion{H}{1} column  density
$\log{N_\mathrm{HI}}  > 18$.  This model is then used as an input to the
SIM\_STIS IDL routine written  by P.  Plait (private  communication), which
accurately models the behavior of STIS.  The output of this routine  is then
convolved by the  expected   PSF and subtracted  from the   observed  spectra.
The residuals   for the     four individual spectra     are   shown  in
Fig.~\ref{fig:excess}.  The S/N of the emission line over the mean of the 4
individual spectra is 23.  This figure also shows for comparison the   spectrum
of  a $6\times10^{-16}~\mbox{ergs} ~ \mbox{s}^{-1}~\mbox{cm}^{-2}$, $\lambda =
2890~\mbox{\AA}$  unresolved emission line convolved by the \textit{HST}+STIS
LSF.  Such a line can indeed be identified as Ly$\alpha$ at $z = 1.38$. 

If due to Ly$\alpha$ at $z=1.38$ this flux would correspond to a Star Formation
Rate (SFR) of about 7 $\mathrm{M_{\sun}\; yr^{-1}}$ following the calculation
in Fynbo et al.\  (2002). This is formally inconsistent with the 2$\sigma$ upper
limit on the SFR of 2.3 $\mathrm{M_{\sun}\; yr^{-1}}$ based on the \ion{O}{3}
line in Vreeswijk, M\o ller \& Fynbo (2003), but the systematic uncertainties
in the relations between Ly$\alpha$ and \ion{O}{3} luminosities and SFR are 
significant enough to leave some room for this interpretation, eventhough we
consider it unlikely. If the line is due to the $z=1.38$
absorber its impact parameter relative to the GRB line-of-sight would have to
be very small, a tenth of an arcsec at most and consistent with zero. Similar
small impact parameters have been observed for DLAs (e.g., M\o ller, Fynbo \&
Fall 2004). Savaglio et al.\ (2002) note that the absorbtion properties of the
$z=1.38$ absorber suggests that it is a DLA.

\subsection{Reddening}
Despite the relatively low \ion{H}{1} column density we detect statistically
significant reddening in the afterglow SED. This is surprising if the dust 
properties
of the ISM are similar to those of SMC dust like for previously studied
GRB afterglows (e.g., Hjorth et al.\ 2003b). For the SMC extinction curve,
for which R$_\mathrm{V} = 2.93$ (Pei 1992), the observed extinction 
corresponds to E(B$-$V)=0.07. For a \ion{H}{1} column 
density of 1--10$\times10^{19}$ cm$^{-2}$ we derive a gas-to-dust
ratio of N(\ion{H}{1})/E(B$-$V) = 1.5--15$\times10^{20}$ cm$^2$ mag$^{-1}$.
This is significantly more reddening per column density than for 
SMC dust, where N(\ion{H}{1})/E(B$-$V) = 4.4$\pm$0.7$\times10^{22}$ cm$^2$ 
mag$^{-1}$ (Bouchet et al.\ 1985). This could imply that a large fraction
of the hydrogen along the line of sight is ionized. Alternatively,
some of the reddening could be due to dust in the two foreground 
\ion{Mg}{2} absorbers, but this appears unlikely as foreground absorbers
in general cause very little reddening (Murphy \& Liske 2004). In other
words, the column densities of the foreground absorbers would have to have
higher column densities than any known DLA,
and this is very unlikely. The intrinsic shape of the afterglow SED is not 
expected to be a pure power-law as the cooling break is located very close 
to the near-UV/optical bands (see below). This may explain some of the 
difference, but we do not expect all the observed bending of the SED
to be due to the intrinsic shape of the SED (see, e.g., Granot \& Sari
2002).

\subsection{Comparison with Afterglow Models}
The post-break decay and spectral slope have been well constrained with 
the data presented in this paper. The spectral slope $\beta$ in the 
optical to near-IR range corrected for extinction is $0.36$ (mean value
of the two measurements in Table~\ref{Table:SED}). The
late-time decay slope $\alpha_2$ is in the range $\alpha_2 = 1.8$--1.9.
These parameters agree well with the results found by H03.
From these afterglow parameters we can infer the position of the 
cooling frequency $\nu_c$ relative to optical frequencies $\nu_\mathrm{O}$ 
in the 
context of the standard blastwave model (see M{\'e}sz{\'a}ros 2002 for 
a review). For $\nu_c > \nu_\mathrm{O}$ we expect
$\alpha_2 - 2 \beta = 1$, and for  $\nu_c < \nu_\mathrm{O}$ we expect
 $\alpha_2 - 2 \beta = 0$ (Sari et al.\ 1999; Chevalier \& Li 1999). Our 
observations 
imply $\alpha_2 - 2 \beta = 1.0$--1.1, clearly indicating that $\nu_c > 
\nu_\mathrm{O}$. This is consistent with the fact that the spectral slope in
the X-rays, $\beta_X = 1.0$, is significantly larger than in the optical,
which makes a spectral break between the optical and X-ray bands 
unavoidable. The change in slope of $\Delta \beta \approx 0.5$ is 
exactly what is expected across the cooling break, $\Delta \beta = 
1/2$. We note that the strong bumps in the optical 
(and possibly X-ray lightcurves, Fox et al.\ 2003) can only 
be interpreted as being the result of density fluctuations if the 
cooling break is bluewards of the optical (and X-rays) (Lazzati et al.\ 
2002, but see also Nakar et al.\ 2003; Nakar \& Piran 2003; 
Bj{\"o}rnsson et al.\ 2004). There is one complication with this 
conclusion: the number $N(E)$ as function of the electron energy 
$E = \gamma_e m_e c^2$ of the electrons producing 
the synchrotron emission is expected to have the form $N(E) \propto E^{-p}$,
and for $\nu_c > \nu_\mathrm{O}$ the theory predicts $\alpha_2 = p$. This 
means that $p \approx 1.9$, which results in a divergent energy spectrum. 
This case has been analyzed by Dai \& Cheng (2001) and Bhattacharya (2001), 
but as $p$ is in our case very close to 2 the resulting relations between 
decay and spectral slopes are quite similar to the equations for the
$p>2$ case.  

The fact that the optical near-IR colors of the afterglow remain
constant means that the cooling break has to be located on the blue side of 
the optical during the period from 0.35 days after the burst and at least 
until October 22 and possibly longer than November 25--26. Depending on
the geometry of the blastwave and the density profile of the surrounding 
medium the cooling break will move towards higher frequencies (wind 
environment,  Chevalier \& Li 1999), lower frequencies (ISM, spherical 
geometry) or remain constant (ISM, jet geometry, Sari et al.\ 1999; 
Chevalier \& Li 1999). 
The fact that the optical to X-ray slope $\beta_\mathrm{OX}$ has remained constant
within the errors from October 7 and until November 26 with a value close to 
$\beta_X$ means that the cooling frequency must have remained constant 
and close to the optical in this time span. This implies a jet geometry and 
a constant density environment. Li \& Chevalier (2003) argue
that an apparent break in the very early lightcurve ($t<0.1\; \mathrm{days}$) 
is best understood assuming a wind-shaped circumburst medium. The only 
way to reconcile this with the discussion above is that the apparent break 
in the optical lightcurve is due to lightcurve fluctuations rather than being 
due to the typical frequency $\nu_m$  passing down from higher frequencies 
as suggested in the wind model of Li \& Chevalier (1999).

The cannonball model offers an alternative explanation for the GRB
phenomenon (Dado, Dar \& De R{\'u}jula 2002). In this model the jet 
opening angle is much smaller,
and the relativistic $\gamma$ factor higher than in the fireball
model. Dado, Dar \& De R{\'u}jula (2003) have presented an analysis of 
GRB~021004 in the cannonball model in which a very good fit to most of 
the available
groundbased data up to about 30 days after the burst was obtained.
In the cannonball model the asymptotic (late-time)  behavior of the 
afterglow is $F_{\nu}(t) \propto t^{-2.13}\nu^{-1.1}$ (Dado et al.\ 2002).
The effective decay slope for the cannonball model fit GRB~021004 
in Dado et al.\ (2003) between October 11 and November 25 is $\alpha_2 = 1.92$,
which is close to the observed value. A spectral slope change
towards $\beta=1.1$ is not observed up to 50 days after the burst.

\subsection{The Host Galaxy}
The host galaxy is extremely blue in the observed optical bands. 
Fig.~\ref{fig:hostenviron} shows the optical colors of the host in
a color-color diagram. The points with error bars represent galaxies
in the environment of the host (from 70$\times$70 arcsec$^2$ around 
the host) and the dots are colors of galaxies from the ACS observations
of the GOODS South field\footnote{For the GOODS data we use the 
average of F775W and F850LP magnitudes as a proxy for F814W}. As seen, 
the host is in the extreme blue color of the distribution for field 
galaxies. The reason for this is a combination of a young age 
for the star burst ($<$100 Myr) 
and the redshift causing a very small Ly$\alpha$ blanketing in the F435W 
filter and placing the Balmer jump beyond the F814W band. The optical 
bands hence all probe the restframe
UV continuum of the newly formed massive stars in the galaxy. A SFR of about
10 M$_{\sun}$ yr$^{-1}$ is derived from the restframe UV flux density. The
SFR inferred from the Ly$\alpha$ luminosity is consistent with this value. 
The host galaxy has been observed in the sub-mm range with SCUBA (Tanvir et al.
2004), but it was not detected above a 2$\sigma$ limit of 2.5 mJy. All
evidence is consistent with the host being a young, dust-poor starburst.
This is typical for GRB host galaxies (Fruchter et al.\ 1999; Le 
Floc'h et al.\ 2003; Christensen et al.\ 2004; Courty, Bj{\"o}rnsson \&
Gudmundsson 2004; Jakobsson et al.\ 2005).  

So far, 15 GRBs have been detected at redshifts where Ly$\alpha$ is observable 
from the ground (see Table~4 in Jakobsson et al.\ 2005). Of these, 
GRB~021004 has the intrinsically brightest detected host galaxy. Nevertheless, 
it is not brighter than the 
characteristic luminosity L$^*$ for Lyman-break galaxies at slightly 
larger redshifts (e.g., Adelberger \& Steidel 2000). The reason why most GRB 
host galaxies are relatively faint, dust-poor starbursts is not yet 
established. It could be that most of star-formation at these redshifts are 
located at the faint end of the luminosity function. There is evidence that 
the faint end slope of the luminosity function at high redshift is 
significantly steeper than in the local Universe (Adelberger \& Steidel 2000; 
Shapley et al.\ 2001), so this is not unlikely (Jakobsson et al.\ 2005). 
However, there is still substantial 
uncertainty about the faint end slope (e.g., Gabasch et al.\ 2004).
Another possibility is a low metallicity preference for GRBs as predicted by 
the collapsar model (MacFadyen \& Woosley 1999; see also Fynbo et al.\ 2003;
Le Floc'h et al.\ 2003, Prochaska et al.\ 2004). With the current very 
inhomogeneous sample of GRB host galaxies we cannot exclude that the current 
sample is somewhat biased against dusty starbursts (see also 	
Ramirez-Ruiz, Trentham \& Blain 2002). The currently operating
Swift mission offers the possibility to resolve this issue (Gehrels et al.\ 
2004). 

\acknowledgments
We thank Arnon Dar and Gunnlaugur Bj{\"o}rnsson for 
critical comments on earlier versions of the manuscript and the
anonymous referee for a thorough and constructive report.
Support for Proposal number GO 9405 was provided by NASA through a 
grant  from the Space Telescope Science Institute, which is operated by 
the Association of  Universities for Research in Astronomy, 
Incorporated, under NASA contract  NAS5-26555. We thank the schedulers of 
the Space Telescope, and in particular our Program Coordinator, Ray Lucas, 
for the extraordinary effort they put in to assure timely observations.
AJL acknowledges support from the Space
Telescope Science Institute Summer Student Programme and from PPARC,
UK. STH acknowledges support from the NASA LTSA grant NAG5--9364. PJ 
acknowledges support from a special grant from the Icelandic Research 
Council.
This work was conducted in part via collaboration  within the the Research and 
Training Network ``Gamma-Ray Bursts: An Enigma and a Tool'', funded by
the European Union under contract number HPRN-CT-2002-00294.
This work was also supported by the Danish Natural Science Research Council 
(SNF), the Carlsberg Foundation, and by the Danish National Research 
Foundation.

\end{document}